\documentstyle[aps,prl,floats,epsfig,preprint]{revtex}
\begin{document}
\tightenlines

\title{Anisotropy of flux-flow resistivity in UPt$_3$}
\author{N. L\"utke-Entrup$^1$, R. Blaauwgeers$^2$, B. Pla\c{c}ais$^1$,
A. Huxley$^3$, S. Kambe$^3$, M. Krusius$^2$, P. Mathieu$^1$ and
Y. Simon$^1$} 
\address{$^1$ Laboratoire de Physique de la Mati\`ere
Condens\'ee de l'ENS, UMR 8551  \footnote{The UMR 8551 is ``Unit\'e Mixte de Recherche'' of
the CNRS, associated with  universities Paris 6 and Paris 7}, 24 rue Lhomond, F-75231 Paris Cedex5,
France \\ 
$^2$ Low
Temperature Laboratory, Helsinki University of Technology, Box 2200,
FIN-02015 HUT, Finland\\ $^3$ D\'epartement de Recherche Fondamentale
sur la Mati\`ere Condens\'ee, CEA, F-38054 Grenoble Cedex5, France}
\date{\today} \maketitle

\begin{abstract}
The ac penetration depth,  $\lambda_{\rm ac}(T,H,f)$, has been
measured in superconducting  UPt$_3$ single crystals for
$\bbox{H}\!\parallel\!\hat c$ and $\bbox{H}\!\perp\!\hat c$ in the
range  $f=$0.01--1\thinspace MHz and $T=$0.1--0.4 K. The
contributions from bulk and  surface pinning have been separated
to yield the flux-flow resistivity $\rho_{\rm f}(H)$. With
$\bbox{H}\!\perp\!\hat c$, $\rho_{\rm f}$ displays
magneto-resistance  at low-field which agrees with previous dc
measurements and the characteristic scaling law of clean crystals
with anisotropic gap. When $\bbox{H}\!\parallel\!\hat c$, the
low-field $\rho_{\rm f}$ is three times larger.   We interpret
this property  as evidence for flux lines with unconventional core
structure.
\end{abstract}
\pacs{PACS numbers: 74.70.Tx, 74.60.Ge, 74.25.Nf, 74.25.Fy.}

The reduced symmetry of the superconducting energy gap in
unconventional superconductors is generally inferred from the
presence of low-energy excitations, which exist due to nodes in
the gap function $\Delta(\bbox{k}_F)$.  In the most important
heavy-fermion compound UPt$_3$, evidence for gap nodes has been
deduced from measurements of the temperature dependences of the
London penetration depth $\lambda(T)$ \cite{Signore95}, ultrasonic
attenuation \cite{Ellman96}, thermal conductivity
\cite{Suderow98},  and most recently flux-flow resistivity
$\rho_{\rm f}(T)$ \cite{Kambe99ff}. UPt$_3$ also has a unique
phase diagram, with two  low-field phases $A$ and $B$ and a
high-field $C$-phase \cite{Fisher89,Adenwalla90} (see insets in
Fig.~\ref{pinning}). These properties are compatible with a
Ginzburg-Landau expansion in terms of a two-component order
parameter \cite{Sigrist91,Sauls94,Lukyanchuk95}. Such states allow
for new and unconventional vortex core structures as well as other
topological defects, such as domain walls or textures
\cite{Sigrist91,Sauls94,Lukyanchuk95,Tokuyasu90}. A multitude of
different vortex structures and other topological defects have
been discovered in $^3$He superfluids \cite{Eltsov}, but so far
direct experimental evidence for their existence in
superconductors is sparse \cite{Kirtley}.

From neutron diffraction measurements it is known that the $B$ and
$C$ phases in UPt$_3$ support distorted lattices of singly
quantized vortex lines, with a clear break in the characteristic
length scales at the $B\rightarrow C$ transition  \cite{Yaron97}.
At low magnetic field parallel to the $c$-axis, the two-component
order-parameter model predicts that three different core
structures of singly quantized vortices are possible in the highly
symmetric $B$-phase \cite{Tokuyasu90}. These have all singular
structure with a radius comparable to the coherence length $\xi$.
One of them is axisymmetric while the other two, called triangular
and crescent cores, are nonaxisymmetric and in contrast to the
former, here the order parameter amplitude does not go to zero
anywhere within the core. The relative stabilities of these
different structures depend on the Ginzburg-Landau parameters,
which are still largely undetermined.

The investigation of the vortex-core structure in the bulk
superconductor requires  measurement of dynamical properties, like
flux-flow dissipation or the Hall angle. We report on measurements
of flux-flow resistivity $\rho_{\rm f}(H,T)$ in the
low-temperature limit ($T\lesssim 0.3$ K) for $\bbox{H}\perp \hat
c$ and $\bbox{H}\parallel \hat c$. For $\bbox{H}\perp \hat c$ our
results are consistent with those of Ref.~\cite{Kambe99ff} at
$T\gtrsim 0.3$ K.  When $\bbox{H}\parallel \hat c$, the
resistivity proves to be much larger than expected. It does not
appear possible to explain this anisotropy unless the vortex
structure is different in the two crystal directions.

Vortex-dynamic measurements are complicated by pinning due to
sample defects, which strongly alter the dc and low-frequency ac
responses \cite{Gittleman66}.  At low temperatures in the
$B$-phase,  the use of large dc currents, to depin vortex lines
and to bias the system into the free-flow state, is prohibited due
to excessive Joule heating. In the present ac measurements this
difficulty is overcome by recording the linear response over two
decades in frequency ($f=0.01$--1\thinspace MHz) while the
dissipation is maintained below $0.1$ nW/mm$^2$. As shown in
Ref.~\cite{Entrup}, the linear regime at low rf excitation level
($\lesssim 1\,\mu$T) and small amplitudes of vortex line
oscillation ($\lesssim 1\,$\AA) lends itself to quantitative
analysis, in spite of strong surface and/or bulk pinning. From the
measurement of the effective ac penetration depth $\lambda_{\rm
ac}(f)$ we extract the damping parameter, the flux-flow
resistivity $\rho_{\rm f}$ or the free-flow skin depth
$\delta_{\rm f}=\sqrt{\rho_{\rm f}/(\pi \mu_0 f)}$. The
frequencies are still sufficiently low so that the anomalous
skin-depth ($\delta_{\rm f}\geq 10\,\mu$m$\gg l_{\rm m}\approx
5000\,$\AA) and relaxation-time effects ($\tau\lesssim 0.2\,$ns
\cite{Kambe99ff}) can be neglected.

Our measurements have been performed on two large single crystals
(labelled B22 and B3b). They were prepared in Grenoble and were
spark cut from the same ingot. Their quality is very similar to
that of the sample used in the earlier dc measurements of
Ref.~\cite{Kambe99ff}. After cutting, the samples were annealed,
but no surface polishing was applied.  Their final dimensions are:
$L(x)\times W(y)\times d(z)=$ $5.5(\hat{a}^*) \times 2.9(\hat{a})
\times  1.16(\hat{c})$ mm$^3$ (B3b) and $5.5(\hat{c}) \times
3.04(\hat{a}^*) \times 0.63(\hat{a})$ mm$^3$ (B22). They have low
residual resistivity, $\rho_{\rm n} = 0.52+1.44 T^2 + 0.02
\,(\mu_0H)^2\; \mu\Omega$cm ($\mu_0H$ in Tesla), for currents
$\bbox{J}\!\perp\!\hat c$, which is the case in our measurements.
In these and similar samples \cite{Suderow98,Kambe99ff}, the
resistivity for $\bbox{J}\!\parallel\!\hat c$ is smaller by a
factor 0.33 -- 0.37 (for $T \lesssim T_{\rm c}$). The measured
$\rho_{\rm n}$ corresponds to a mean-free path $l_{\rm m}\gtrsim
10 \,\xi$, proving that the samples represent the moderately
\emph{clean limit} \cite{Kycia98}.

The dc field $\mu_0 H\leq 3 \,$T and the vortex lines are along
the $\hat z$ direction.  The excitation field, $h \,
\mathrm{e}^{-{\mathrm i}2\pi f t}$, is applied along the $\hat x$
direction so that the vortices oscillate in the $xz$ plane, and
currents and electric fields $\bbox{E}$ are induced in the $\hat
y$ direction. Due to the Lorentz-force anisotropy, $E_y$ is
concentrated on the two main faces $z=\pm d/2$ so that the sample
mimics the 1-dim geometry of an  infinite slab.  The measured
signal is the flux $\Phi_{\rm ac}$ through a 15-turn pick-up coil
wound directly around the sample in the $\hat x$ direction.  The
apparent complex penetration depth is defined as $\lambda_{\rm ac}
=  \lambda'+{\mathrm i} \lambda'' = [\Phi_{\rm
ac}(H)\!-\!\Phi_{\rm ac}(0)]/(2\mu_0hW)$.

The sample with its pick-up coil and the slightly larger
excitation solenoid are placed inside the mixing chamber of a
$^3$He-$^4$He dilution refrigerator. The measurements are
performed in the temperature range $T = 0.1$--$0.4\,$K, by
recording the spectrum $\lambda_{\rm ac}(f)$ at fixed field $H$
and by moving from one field value to the next, while the
temperature is kept constant with a feed-back loop. The
temperature is monitored with a calibrated Ge resistance
thermometer inside the mixing chamber. Its field dependence is
adjusted by comparing the measured $H_{c2}$ values with those in
Ref.~\cite{Adenwalla90}. The amplitude and phase of the signal
voltage are calibrated against the difference between the
responses in the normal and the Meissner states. This
normalization procedure yields a phase accuracy better than
$1^{\circ}$ and a resolution $\delta\lambda_{\rm ac}\lesssim1 \,
\mu$m \cite{Entrup}.

Typical  spectra of $\lambda_{\rm ac}(f)$ are shown in
Fig.~\ref{spectra}. The complete data set consists of more than
200 different spectra. The real and imaginary parts of
$\lambda_{\rm ac}$ can  be accurately fit with the following
formula \cite{Entrup,Entrup2000}
\begin{equation}
\frac{1}{\lambda_{\rm ac}}=\frac{1}{L_{\rm S}}+
\sqrt{\frac{1}{\lambda_{\rm C}^2}-\frac{2{\mathrm i}}{\delta_{\rm
f}^2}}\;\; . \label{spectrum}
\end{equation}
$L_{\rm S}(H,T)$ and $\lambda_{\rm C}(H,T)$ are two
frequency-independent lengths, which describe surface and bulk
pinning respectively.  The high-frequency limit,  $\lambda_{\rm
ac} (f \!\rightarrow\! \infty) \!=\! (1+\mathrm i)\delta_{\rm
f}/2$, corresponds to the ideal flux-flow  response, while the
low-frequency limit, the quasistatic response, $\lambda_{\rm ac}
(f \!\rightarrow \! 0) \!=\! \lambda'(0)\!=\!(\lambda_{\rm C}^{-1}
+ L_{\rm S}^{-1})^{-1}$, is purely inductive and cannot
discriminate between surface and bulk pinning. The relative weight
of surface and bulk pinning becomes apparent in the crossover
regime.

The excitation field penetrates as the sum of two modes. The first
mode, localized near the surface, is associated with strong
screening currents, whose amplitudes are governed by surface
roughness. If the bulk sample is free of defects, the screened
field penetrates further over the free-flow depth $\delta_{\rm
f}$. This situation ($\lambda_{\rm C}\! \rightarrow \! \infty$)
has been systematically the case in the conventional
superconductors, which so far have been measured \cite{Entrup}. In
contrast, if there are bulk defects, such as those usually
introduced in classical theories of pinning
\cite{Gittleman66}, the bulk mode is strongly
attenuated, but penetrates at low frequencies over a Campbell
length $\lambda_{\rm C}$ ($\!\ll\!\delta_{\rm f}$ at low
frequencies).

Curiously enough, our moderately clean UPt$_3$ crystals display a
large bulk pinning strength $1/\lambda_{\rm C}$, as evident from
Fig.~\ref{pinning}. The data in Fig.~\ref{pinning} show that in
both crystal orientations at low fields pinning is strong and
dominated by the surface process. This is not surprising, since
after spark cutting the surfaces are visibly rough. With
increasing field, surface pinning falls down and finally vanishes
at a reversible field value close below the $B \!\rightarrow \! C$
transition. This result is consistent with recent observations in
point contact spectroscopy \cite{DeWilde94,Goll95,Obermair98},
which suggest a suppression of the order parameter at the surface
at higher fields. Bulk pinning decreases more slowly as a function
of ($H_{\rm c2}\!-\!H$) and is the dominant source of pinning in
the $C$ phase. No clear anomaly can be distinguished at the $B \!
\rightarrow \! C$ transition \cite{Entrup2000}. Nor do we observe
hysteresis as a function of the field-sweep direction. Low-field
vortex-creep measurements \cite{Amann96} have suggested that new
mechanisms could be present, such as intrinsic pinning by domain
walls in the bulk \cite{Sigrist91}. If this is the source for the
bulk pinning, then the domain walls have to persist in the $C$
phase. Additional work, including variations in surface treatment,
should provide better insight in the pinning processes.

The normalized flux-flow resistivity, $\rho_{\rm f}/\rho_{\rm n}$,
is shown in Fig.~\ref{rhoF}. As expected for the moderately clean
superconductor, $\rho_{\rm f}(H)$ exceeds the ``normal-core
limit'', $\rho_{\rm f}\!=\!\rho_{\rm n}H/H_{\rm c2}$, in both
crystal directions. When the contributions from bulk and surface
pinning have been removed, the $\rho_{\rm f}(H)$ data become
sample independent, as it should be. In the direction
$\bbox{H}\!\perp\!\hat c$, where comparison is possible,
$\rho_{\rm f}(H)$ agrees quantitatively with the earlier dc
measurements of Ref.~\cite{Kambe99ff}. It is usual to define the
initial slope in Fig.~\ref{rhoF} as
\begin{equation}
\frac{\rho_{\rm f}(H)}{\rho_{\rm n}(H)}= r_0 \frac{H}{H_{\rm
c2}}\;. \label{magnetoresitance}
\end{equation}
At low-field the scaling factor $r_0(T)$ obtains, in the
perpendicular direction, the value $r_{0\perp}\!=\!1.6\pm0.15$ in
the low-temperature limit. In fact, as seen in Fig.~\ref{rhoF},
below $0.3\,$K$\, \simeq 0.6 \; T_{\rm c}$, $r_0$ is already
temperature independent. This is in agreement with the results in
Ref.~\cite{Kambe99ff}, when they are extrapolated to low
temperatures, and testifies for good general consistency between
two different measuring methods and samples. It should be noted
that the temperature independence below 0.3 K also embraces the
bulk and surface pinning strengths in Fig.~\ref{pinning}.

In effect, Eq.~(\ref{magnetoresitance}) states that, at low field,
flux-flow dissipation is additive and therefore proportional to
the vortex density, $n=\mu_0 H/\varphi_0$. Assuming that we can
identify $\varphi_0/\mu_0H_{c2}$ with $2\pi b^2$, where  $b$ is an
effective core radius, as is the case for conventional Abrikosov
vortices, then Eq.~(\ref{magnetoresitance}) can be written as
\begin{equation}
\frac{\rho_{\rm f}(H)}{\rho_{\rm n}(H)}= r^* 2\pi n b^2 \; .
\label{core_resistivity}
\end{equation}
Here $r^*\! = \! r_0$ for Abrikosov vortices. More generally, when
$H_{\rm c2}$ is not an appropriate scaling parameter due to Pauli
paramagnetic limitation of $H_{\rm c2}$ and/or possible changes in
the core structure, $r^*(T)$ can still be considered a
dimensionless factor which depends on the mechanism of
dissipation.

In the dirty limit ($l_{\rm m}\ll\xi$) one has $r_0\lesssim1$ and
Eq.~(\ref{magnetoresitance}) reads $\rho_{\rm f}/\rho_{\rm n}\approx
H/H_{c2}$ (normal-core model).  In the clean limit dissipation is
governed by the relaxation of the quasiparticle excitations localized
in the vortex core. We shall make use of the theory by Kopnin \emph{et
al.} \cite{Kopnin95}. They take $b\!=\!\xi$ as the size of the
confining vortex-core potential, allow for anisotropy in
$\Delta(\bbox{k}_F)$ over a spherical Fermi surface, and predict that
\begin{equation}
r^*(T)\simeq\alpha \frac{k_{\rm B}T_{\rm c}}{\Delta_{\rm
max}(T)}\;,\quad \alpha=\frac{\frac{2}{3}\Delta_{\rm max}^2}{\langle
(1\!-\!(\hat{k}\cdot\hat{H})^2)\Delta^2(\bbox{k})\rangle} \;.
\label{Kopnin}
\end{equation}
The $\langle \rangle$ brackets denote a weighted average over the
Fermi surface. For an isotropic gap (s-wave superconductor) the
dimensionless factor $\alpha$ is unity. If $\Delta(\bbox{k})$ has
nodal structure, or at least strong anisotropy, then $\alpha$ is
expected to be larger than unity and its value depends on the
orientation of the vortex lines with respect to the crystal axes.  If
we take $\Delta_{\rm max}(0)\!=\! 1.9 \, k_{\rm B}T_{\rm c}$
\cite{DeWilde94} (and $r^*\!=\!r_0$), Eq.~(\ref{Kopnin}) yields
$\alpha_\perp \! = \!  1.6\!\times \! 1.9 \!\simeq \!3.0$, which
agrees within the combined experimental precisions with the value 3.2
obtained in Ref.~\cite{Kambe99ff}. The large $\alpha$ is as expected
because of gap anisotropy.

The analysis of the measurements in terms of Eq.~(\ref{Kopnin})
applies for conventional superconductors with $r_0\!=\!r^*$, or
equivalently, $b_\perp^2\!= \!\xi_a\xi_c\!= \!\varphi_0/
(2\pi\mu_0H_{c2\perp})$. Here $H_{c2\parallel}$ is not a good scaling
parameter because of the Pauli paramagnetic limitation at low
temperature in this direction. To compare the results in the two
orientations, we use Eq.~(\ref{core_resistivity}) by taking
$b_{\perp}^2 \!=\!  \xi_{a}\xi_{c}$ and $b_{\parallel}^2\!
=\!\xi_{a}^2$. Furthermore, taking the high-temperature anisotropy
$\xi_{a}\!=\!0.6\xi_{c}$, we find the low-temperature values of $r^*$
to be $r^*_{\perp} \!=\!1.6$ and $r^*_{\parallel}
\!=\!4.7/0.6\!=\!7.8$. Eq.~(\ref{Kopnin}) gives that these correspond
to $\alpha_{\perp} \!=\!3.0$ and $\alpha_{\parallel} \!= \!7.8\times
1.9 = 14.8$, which represents a large anisotropy of
$\alpha_\parallel/\alpha_\perp\!\simeq\!5$.

An anisotropy in $\alpha$ as large as this cannot be explained by
the UPt$_3$ structure, by taking into account the possible
anisotropies from the $D_{6h}$ point group symmetry in the
expression of $\alpha$ in Eq.~(\ref{Kopnin}).  For instance, let
us consider the gap structure in the 2-dim representations
$E_{1g}$ ($\Delta\!\sim\!\{k_zk_x,k_zk_y\}$) and $E_{2u}$
($\Delta\!\sim\!\{k_z(k_x^2-k_y^2),2k_zk_xk_y\}$), which are
appropriate to the $A$ and $B$ phases  \cite{Sauls94}. For
$E_{1g}$ ($\Delta\!\sim\!k_z(k_x\!\pm\!\mathrm{i}k_y)$, $B$-phase)
we find $\alpha_\parallel\! =\!1.25 \, \alpha_\perp\!= \!4.4$. For
$E_{2u}$, which is strongly supported by experiment
\cite{Ellman96,Graf2000}, there is no anisotropy at all:
$\alpha_\parallel\! =\!\alpha_\perp\! =\!3.9$.  Thus the
small jump of about $10\%$ in $\rho_{\rm f}/\rho_{\rm n}$, which was
observed in Ref.~\cite{Kambe99ff} at the $A\!\rightarrow\!B$
transition for vortices in the $aa^*$ plane, might be ascribed
directly to anisotropy in crystal structure and symmetry breaking, but
not the large value of $r_{0 \parallel}/r_{0 \perp}$.

Quasiparticle scattering may additionally contribute to the
anisotropy in $\rho_{\rm f}(\bbox{H})$. An estimate can be worked
out by comparing the anisotropy in $\rho_{\rm n}(\bbox{J})$ to
that in the slope of the critical field $\bbox{H}_{\rm c2}(T)$ at
$T_{\rm c}$: $dH_{\rm c2,c}/dT \simeq -7.2$ T/K and $dH_{\rm
c2,a}/dT \simeq -4.6$ T/K  \cite{Suderow98}. These values yield
for the ratio of the coherence lengths $\gamma = \xi_{\rm
a}/\xi_{\rm c} \simeq 0.64$, which corresponds to an effective
mass ratio of $m^*_{\rm c}/m^*_{\rm a} = \gamma^2 \simeq 0.41$.
The anisotropy in $\rho_{\rm n}(\bbox{J})$ amounts to a ratio
$\rho_{\rm n,c}/\rho_{\rm n,a} \simeq$ 0.33 -- 0.37. This value
includes the anisotropies in effective masses and in quasiparticle
relaxation times. Comparing the two estimates we conclude that not
much can be allowed for the anisotropy in quasiparticle
scattering.

Finally, one more source for anisotropy is a different vortex core
structure in the two orientations. The 2-dim Ginzburg-Landau
calculations in Ref.~\cite{Tokuyasu90} suggest the existence of
several competing core structures in the axial direction $\bbox{H}
\!\parallel \!\hat c$ and thus the perpendicular direction is
likely to be again different. In Fig.~\ref{rhoF} the large
$\rho_{\rm f}/\rho_{\rm n}$ measured with
$\bbox{H}\!\parallel\!\hat c$ could then be explained by assuming
$r^*_\parallel\!=\!r^*_\perp$ in Eq.~(\ref{core_resistivity}) and
$b_\parallel^2 = 3\, b_{\perp}^2 = 3 \xi_a\xi_c$ rather than
$b_\parallel^2 \!=\!\xi_{a}^2$, as was done above. If we ascribe
in this way the entire observed anisotropy to an increase in the
effective core radius in the axial orientation and take
$b_{\parallel}\!=\!\tilde\xi$, we estimate the ratio of the fourth
order Ginzburg-Landau coefficients to be $\beta_2/\beta_1\!
=\!\xi_a^2/ \tilde\xi^2\!\gtrsim \!0.2$. This is in fair
agreement with the independent determination $\beta_2/\beta_1\!
\sim\!0.2$--$0.5$ extracted from the specific-heat jump at $T_{\rm
c}$ \cite{Fisher89,Sauls94}.

Such an interpretation assumes that the effective core  radius in
the axial orientation acquires the value $\tilde\xi \approx
\sqrt{3 \xi_a\xi_c} \approx 2\xi_a$. This is possible if the
singular ``hard vortex core'' has reduced rotational symmetry, as
is the case in $^3$He-B at low temperatures \cite{Eltsov}. A
second possibility is that the singular hard core lies embedded
within a larger ``soft-core'' which has an effective radius
comparable to $\tilde\xi$ \cite{Eltsov}. In summary, our
measurements provide new indication for the presence of
unconventional vortex structures in UPt$_3$.

We acknowledge instructive discussions with N. Kopnin and J.
Flouquet. N.L-E. and B.P.  thank the LTL for hospitality. This
work was funded by the EU Training and Mobility of Researchers
Programme (contract no.  ERBFMGETC980122).

\newpage
\begin{figure}[!!!t]
\caption{Spectra of the apparent penetration depth $\lambda_{\rm
ac}(f)=$ $ \lambda'+\mathrm{i}\lambda''$, showing as a function of
frequency the crossover from  pinning-dominated  to  free-flow
response ({\it ie.} the high frequency plateau
$\lambda''\!\simeq\!\lambda'\!=\!\delta_{\rm f}/2$). The {\it
solid curves} are  fits to Eq.~(\protect\ref{spectrum}) with
$L_{\rm S}$, $\lambda_{\rm C}$, and $\rho_{\rm f}$ as adjustable
parameters. To illustrate their relative influence on the fitting,
the {\it dashed curves} represent pure bulk-pinning $(1/L_{\rm
S}=0)$ and the {\it dash-dotted} curves pure surface-pinning
$(1/\lambda_{\rm C}=0)$.} \label{spectra}
\end{figure}

\begin{figure}[!!!t]
\caption{Field dependence of the bulk and surface pinning
strengths, $1/\lambda_{\rm C}$ {\it (full symbols)} and $1/L_{\rm
S}$ {\it (open symbols)} in two crystal orientations. The
different symbols denote measurements at different temperatures
below 0.33 K. The data are derived from fits to the measured
$\lambda_{\rm ac}(f)$ spectra, as shown in
Fig.~\protect\ref{spectra}. Bulk pinning $1/\lambda_{\rm C}$
vanishes at $H_{\rm c2}$, while the surface contribution $1/L_{\rm
S}$ decreases more rapidly and vanishes just below the $B
\!\rightarrow \! C$ transition. The surface-pinning fraction is
shown in gray-scale units in the {\it inset}, together with the
UPt$_3$ phase diagram as a function of field and temperature.}
\label{pinning}
\end{figure}

\begin{figure}[!!!tt]
\caption{Flux-flow resistivity $\rho_{\rm f}(H)$, as deduced by
fitting the measured $\lambda_{\rm ac}(f)$ (Fig.~\ref{spectra}) to
Eq.~(\protect{\ref{spectrum}}). It is seen that  $\rho_{\rm f}(H)$
becomes temperature independent below $0.3\,$K as a function of
$H/H_{c2}$.  The results differ when measured with
$\bbox{H}\parallel\hat{c}$ (sample B3b) or $\bbox{H}\perp\hat{c}$
(sample B22). At low fields this anisotropy is quantified by the
initial slopes $r_0(T)$ in Eq.~(\ref{magnetoresitance}) ({\it
dashed lines}), yielding $r_{0\perp}\!=\!1.6\pm0.15$ and
$r_{0\parallel}\!=\!4.7\pm0.3$.  The inset shows unscaled data at
$0.15\,$K such that $H_{c2}$ coincides in the two directions.  The
overshoot in $\rho_{\rm f}$, just below $H_{\rm c2}$, is an
artifact due to the onset of flux penetration at the edges of the
sample. In the main frame as well as in Fig.~\protect\ref{pinning}
such data points have been omitted. } \label{rhoF}
\end{figure}

\begin{thebibliography}{99}

\bibitem{Signore95} P.J.C. Signore \emph{et al.},  Phys. Rev. B {\bf
52}, 4446 (1995).

\bibitem{Ellman96} B. Ellman \emph{et al.},
Phys. Rev. B {\bf 54}, 9043 (1996).

\bibitem{Suderow98} H. Suderow \emph{et al.},  Phys. Rev. Lett. {\bf
80}, 165 (1998).

\bibitem{Kambe99ff} S. Kambe \emph{et al.}, Phys. Rev. Lett. {\bf 83},
1842 (1999).

\bibitem{Fisher89} R.A. Fisher \emph{et al.}, Phys. Rev. Lett. {\bf
62}, 1411 (1989).

\bibitem{Adenwalla90} A. Adenwalla \emph{et al.},
Phys. Rev. Lett. {\bf 65}, 2298 (1990).

\bibitem{Sigrist91} M. Sigrist and K. Ueda, Rev. Mod. Phys. {\bf 63},
239 (1991).

\bibitem{Sauls94} J.A. Sauls, Adv. Phys.  {\bf 43}, 113 (1994).

\bibitem{Lukyanchuk95} I.A. Luk'yanchuk and M.E. Zhitomirsky,
Supercond. Rev.  {\bf 1}, 207 (1995).

\bibitem{Tokuyasu90} T. Tokuyasu \emph{et al.}, Phys. Rev. B {\bf 41},
8891 (1990).

\bibitem{Eltsov} V.B. Eltsov and M. Krusius, in {\it Topological defects in
$^3$He superfluids and the non-equilibrium dynamics of
symmetry-breaking phase transitions}, eds. Yu. Bunkov and H.
Godfrin (Kluwer Academic Publ., Dordrecht, 2000), p. 325.

\bibitem{Kirtley} J.R. Kirtley {\it et al.}, Phys. Rev. Lett. {\bf
76}, 1336 (1996).

\bibitem{Yaron97} U. Yaron \emph{et al.}, Phys. Rev. Lett. {\bf 78},
3185 (1997).

\bibitem{Gittleman66}J.I. Gittleman and B. Rosenblum,
Phys. Rev. Lett. {\bf 16}, 734 (1966); A.M. Campbell, J. Phys. C {\bf 2}, 1492
(1969).

\bibitem{Entrup} N. L\"utke-Entrup \emph{et al.}, Phys. Rev. Lett. {\bf
79}, 2538 (1997); Physica B {\bf 255}, 75 (1998).

\bibitem{Kycia98} J.B. Kycia \emph{et al.}, Phys. Rev. B {\bf 58},
R603 (1998).

\bibitem{Entrup2000} N. L\"utke-Entrup \emph{et al.}, Physica B {\bf
284--288}, Pt. II, \emph{ in press} (2000).

\bibitem{DeWilde94} Y. De Wilde \emph{et al.}, Phys. Rev. Lett. {\bf
72}, 2278 (1994).

\bibitem{Goll95} G. Goll \emph{et al.},  Phys. Rev. B {\bf 52}, 6801
(1995).

\bibitem{Obermair98} C. Obermair \emph{et al.}, Phys. Rev. B {\bf 57},
7506 (1998).

\bibitem{Amann96} A. Amann  \emph{et al.}, Europhys. Lett.  {\bf
33}, 303 (1996).

\bibitem{Kopnin95} N.B. Kopnin and A.V. Lopatin, Phys. Rev. B {\bf
51}, 15291 (1995); J. Low Temp. Phys. {\bf 110}, 885 (1998).

\bibitem{Graf2000} M.J. Graf  \emph{et al.},
 Physica B {\bf 280}, 176 (2000).

\end{thebibliography}
\end{document}